\documentclass[aps,pra,groupedaddress,superscriptaddress,twocolumn,10pt]{revtex4-1}
\usepackage[utf8]{inputenc}
\usepackage{hyperref}
\usepackage{graphicx}
\usepackage{setspace}

\usepackage{siunitx}
\usepackage{braket}
\usepackage{amssymb}
\usepackage{amsmath}

\begin{document}

\title{Experimental realization of a 3D long-range random hopping model}

\author{Carsten Lippe}
\affiliation{Department of Physics and Research Center OPTIMAS, Technische Universität Kaiserslautern, 67663 Kaiserslautern, Germany}

\author{Tanita Klas}
\affiliation{Department of Physics and Research Center OPTIMAS, Technische Universität Kaiserslautern, 67663 Kaiserslautern, Germany}

\author{Jana Bender}
\affiliation{Department of Physics and Research Center OPTIMAS, Technische Universität Kaiserslautern, 67663 Kaiserslautern, Germany}

\author{Patrick Mischke}
\affiliation{Department of Physics and Research Center OPTIMAS, Technische Universität Kaiserslautern, 67663 Kaiserslautern, Germany}

\author{Thomas Niederprüm}
\affiliation{Department of Physics and Research Center OPTIMAS, Technische Universität Kaiserslautern, 67663 Kaiserslautern, Germany}

\author{Herwig Ott}
\email{ott@physik.uni-kl.de}
\affiliation{Department of Physics and Research Center OPTIMAS, Technische Universität Kaiserslautern, 67663 Kaiserslautern, Germany}

\date{\today}

\begin{abstract}
Randomness and disorder have strong impact on transport processes in quantum systems and give rise to phenomena such as Anderson localization \cite{Anderson1958,Aspect2008,Modugno2015}, many-body localization \cite{Abanin2019} or glassy dynamics \cite{Bender1986}.
Their characteristics thereby depend on the strength and type of disorder.
An important class are hopping models, where particles or excitations move through a system which has randomized couplings.
This includes, e.g., spin glasses \cite{Bender1986}, coupled optical waveguides \cite{Martin2011}, or NV center arrays \cite{Hanson2008}. They are also key to understand excitation transport in molecular and biological systems, such as light harvesting complexes \cite{Engel2007}.
In many of those systems, the microscopic coupling mechanism is provided by the dipole--dipole interaction.
Rydberg systems \cite{Jones_2017} are therefore a natural candidate to study random hopping models.
Here, we experimentally study a three-dimensional many-body Rydberg system with random dipole--dipole couplings.
We measure the spectrum of the many-body system and find good agreement with an effective spin model.
We also find spectroscopic signatures of a localization--delocalization transition.
Our results pave the way to study transport processes and localization phenomena in random hopping models in detail.
The inclusion of strong correlations is experimentally straightforward and will allow to study the interplay between random hopping and localization in strongly interacting systems.
\end{abstract}

\maketitle

For the description of many-body dynamics in various physical systems, ranging from condensed matter systems over ultracold gases to elementary particles, spin models have proven very beneficial.
A model particularly relevant to particle and energy transport is the XY model \cite{Lieb1961} that describes coupled two-dimensional spin-$\frac{1}{2}$ particles by the Hamiltonian
\begin{equation}\label{eq:XYModel}
    \hat{H}_{XY} = \sum_{i<j}^n{\frac{J_{ij}}{2} \left(\hat{\sigma}_i^x\hat{\sigma}_j^x + \hat{\sigma}_i^y \hat{\sigma}_j^y\right)} + \sum_{i=1}^n E_i \hat{\sigma}_i^z,
\end{equation}
where $\hat{\sigma}_i^{x,y,z}$ denote the Pauli matrices and $J_{ij}$ describes the coupling between spins $i$ and $j$.
Rewriting the term $\left(\hat{\sigma}_i^x\hat{\sigma}_j^x + \hat{\sigma}_i^y \hat{\sigma}_j^y\right)/2 = \hat{\sigma}_i^+\hat{\sigma}_j^- + \hat{\sigma}_i^- \hat{\sigma}_j^+$ by the ladder operators emphasizes the hopping character of the XY model.
The second term describes the on-site energy $E_i$.

For a realistic description of transport processes in solids, the inclusion of defects and disorder was found to be crucial. This way, the metal insulator transition could be explained by assuming constant nearest-neighbor coupling $J_{i,i+1}=J$ and disordered longitudinal fields $E_i$ \cite{Anderson1958,Garanin2013}. This is commonly referred to as the Anderson model.
Several extensions have been considered since then. 
Most prominently, the inclusion of on-site interactions has lead to the rapidly growing field of many-body localization \cite{Abanin2019,}.
Also models with random on-site energy and long-range interaction were found to show many-body localized states \cite{Yao2014,Burin2015}.
In many realistic physical systems, interactions and transport processes are governed by power-laws. 
Thus, in case of position disorder, hopping disorder emerges naturally and results in a randomized coupling strength $J_{ij}$. 
It was found that such random hopping processes lead to similar localization \cite{Eilmes1998}. 

Here, we are interested in hopping disorder mediated by power-law interactions. Such models have received broad interest due to a large variety of emerging physical effects such as many-body relaxation dynamics \cite{Orioli2018}, glassy dynamics \cite{Signoles2019}, localization phenomena \cite{Scholak2014,Marcuzzi2017,Abumwis2020} or superfluid stiffness \cite{Maccari2019}.
Due to their long-range interactions, Rydberg gases are particularly well suited to this purpose.
For atoms prepared in different dipole-coupled Rydberg states $\ket{\mathrm{S}}$ ($ \equiv\ket{\downarrow}$) and $\ket{\mathrm{P}}$ ($\equiv \ket{\uparrow}$), the anisotropic dipole--dipole interaction $\hat{V}^{dd}_{ij} = (\hat{\boldsymbol{d}}_i \cdot \hat{\boldsymbol{d}}_j - 3(\hat{\boldsymbol{d}}_i \cdot \boldsymbol{e}_R)(\hat{\boldsymbol{d}}_j \cdot \boldsymbol{e}_R))/R_{ij}^3$ realizes the XY spin-exchange term of Eq.\,\eqref{eq:XYModel} \cite{Barredo2015} and the couplings read $J_{ij} = C_3 (1-3\cos^2{\theta_{ij}})/R_{ij}^3$, where $R_{ij}$ is the distance between the randomly placed atoms $i$ and $j$ and $\theta_{ij}$ the angle between the quantization axis and their interparticle axis.
The system is governed by the Hamiltonian

\begin{equation}\label{eq:Hamiltonian}
    \hat{H} = \sum_{i<j}^n{\frac{C_3\left( \theta\right)}{R_{ij}^3} \left(\hat{\sigma}_i^+\hat{\sigma}_j^- + \hat{\sigma}_i^- \hat{\sigma}_j^+\right)} + \sum_{\nu=\downarrow,\uparrow}\sum_{i<j}^n\frac{C_6^\nu(\theta)}{R_{ij}^6}\hat{n}_i^\nu\hat{n}_j^\nu,
\end{equation}

where $\hat{n}_i^{\downarrow/\uparrow} = (1 \pm \hat{\sigma}_i^z)/2$ counts the number of $\ket{\downarrow}$ and $\ket{\uparrow}$-excitations. To a much lesser extent, the Rydberg system also realizes an Ising-type term through the interaction between two identical spins $\ket{\downarrow\downarrow}$ or $\ket{\uparrow\uparrow}$ by means of the van der Waals interaction, i.e. $U_{ij} = C_6(\theta)/R_{ij}^6$ \cite{Leseleuc2018a}.
This correction is described by the second term in Eq.\,\eqref{eq:Hamiltonian}.

\begin{figure}[t]
    \includegraphics[width=\columnwidth]{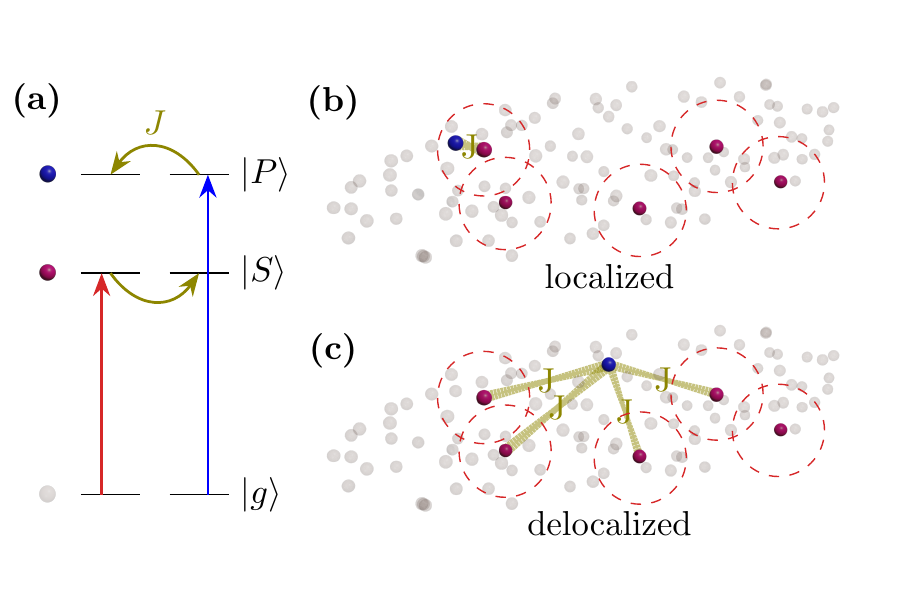}
    \caption{Sketch of the experiment.
    (a) Dipole--dipole interaction between two atoms. One ground state atom $\ket{g}$ (gray) is excited to a Rydberg $\ket{\mathrm{S}}$-state (red) with a two-photon transition (red arrow). A second atom is excited to a Rydberg $\ket{\mathrm{P}}$-state (blue) with a single-photon transition (blue arrow).
    Hopping with strength $J$ (yellow arrows) is induced by resonant dipole--dipole coupling between the two Rydberg states of opposite parity.
    (b) and (c) Spatial distribution of Rydberg excitations corresponding to localized  and delocalized states. Seed atoms in the Rydberg $\ket{\mathrm{S}}$-state (red) are separated by the Rydberg blockade radius (red dashed circles). Probe excitations to the Rydberg $\ket{\mathrm{P}}$-state are shown in blue. Surrounding ground state atoms are shown in gray.
    Yellow connections illustrate the strongest hopping contributions. They can be restricted to two sites only, forming a localized dimer state (b), or to multiple similarly spaced sites, forming a delocalized state (c).}
    \label{fig:Fig1}
\end{figure}

The experiments are carried out in a three-dimensional frozen Rydberg gas without underlying regular lattice structure (Fig.\,\ref{fig:Fig1}).
The $R^{-3}$ scaling thereby ensures that one atom is coupled to many others. The couplings themselves are not purely random, as the atomic arrangement as well as the Rydberg blockade impose inherent correlations on the position disorder due to the triangular inequality and the distance constraints.
In the experimentally realized limit of weak probing the $C_6$ term of Eq.\,\eqref{eq:Hamiltonian} effectively simplifies to a random field $\sum_{i=1}^n E_i\hat{\sigma}_i^z$ and the Hamiltonian takes the form of a pure XY model Eq.\,\eqref{eq:XYModel}.
Note that with a proper choice of the involved Rydberg states, the relative strength of the two terms in Eq.\,\eqref{eq:Hamiltonian} can be tuned with respect to each other, thus allowing to study the full crossover from hopping disorder to on-site disorder. 

The question of localization in dipole--dipole interacting Rydberg systems is subtle. For power-law depending hopping models in a cubic lattice with random on-site energy, a critical dimension analysis reveals a breakdown of localization in three dimensions \cite{Burin2015}, induced by the hopping. Recent studies show, however, that the addition of hopping disorder can restore localization \cite{Scholak2014, Abumwis2020}.
In fact, the eigenstates are expected to show a transition from a regime with predominantly delocalized states to pair-localized states \cite{Scholak2014}, depending on the energy of the state (Fig.\,\ref{fig:Fig1}). 
We here study the Hamiltonian Eq.\,\eqref{eq:Hamiltonian} spectroscopically. This gives access to the density of states of the many-body system and its scaling properties. This approach allows for a direct comparison with numerical simulations and helps us to identify signatures for the appearance of localized and delocalized states.

\section*{Experimental realization}
\begin{figure*}[t]
    \begin{center}
        \includegraphics[width=\textwidth]{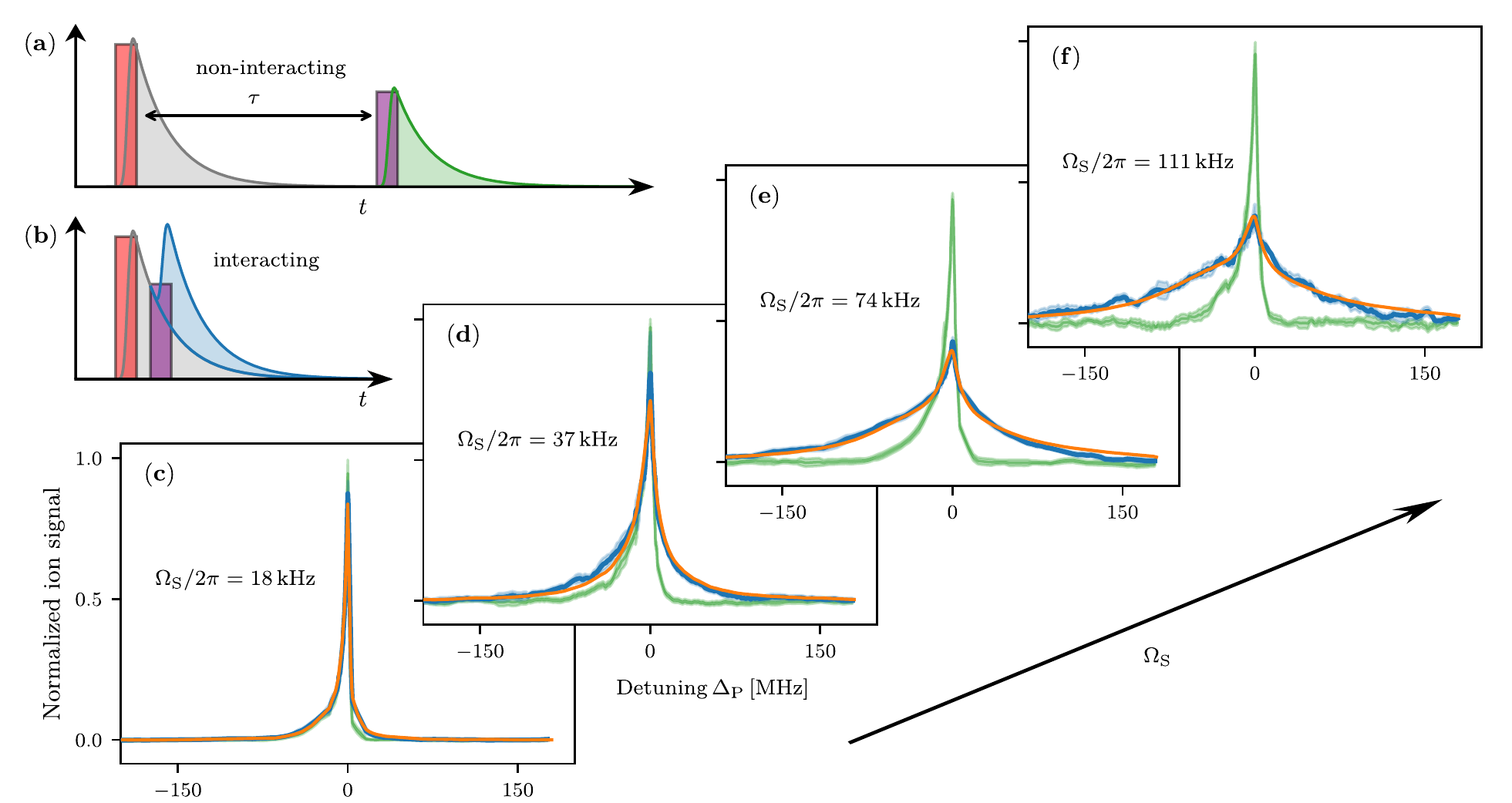}
    \end{center}
    \caption{Spectroscopy of dipole--dipole coupled many-body Rydberg systems.
        (a) and (b) illustrate the pump--probe excitation scheme with delay $\tau$ in the non-interacting (a) and interacting (b) case. The first pulse (red) creates seed excitations in the $\ket{51\mathrm{S}_{1/2}}$-state. The second pulse (purple) excites atoms to the $\ket{51\mathrm{P}_{3/2}}$-state.
        (c) - (f) Spectra for the excitation of the $\ket{51\mathrm{P}_{3/2}}$-state after the creation of different numbers of initial seeds in the $\ket{51\mathrm{S}_{1/2}}$-state with two-photon Rabi frequencies $\Omega_\mathrm{S}$ (c) $2\pi \times \SI{18}{kHz}$, (d) $2\pi \times \SI{37}{kHz}$, (e) $2\pi \times \SI{74}{kHz}$, (f) $2\pi \times \SI{111}{kHz}$.
        The interacting ($\tau=\SI{1}{\mu s}$, blue) and non-interacting ($\tau=\SI{300}{\mu s}$, green) spectra in (c)-(f) are obtained by integrating the blue and green shaded area of the time-resolved signals in (a) and (b) for each detuning $\Delta_\mathrm{P}$, respectively.
        In the interacting case (b), the signal from the probe pulse (blue shaded area) is isolated by subtracting the pump signal (gray shaded area) from the non-interacting measurement (a).
        The results of the numerical model are shown as orange lines.
        We extract average seed excitation numbers $\bar{n}$ of (c) 0.5, (d) 2.1, (e) 4.8 and (f) 5.4.
    }
    \label{fig:Fig2}
\end{figure*}

The experimental realization requires a gas with high number density in order to realize also very small interatomic distances such that localized pair states can be excited.
To this purpose, we prepare a Bose-Einstein condensate (BEC) of $^{87}\mathrm{Rb}$ with an atom number of $\num{90e3}$ and a peak density of $\SI{3e14}{cm^{-3}}$.
The sample is spin-polarized in the $\ket{5\mathrm{S}_{1/2}, F=2, m_F=2}$ ground state.
The experiment is carried out in a pump--probe scheme (Fig.\,\ref{fig:Fig2}a-b), where in a first step a variable number of atoms is brought into the $\ket{\downarrow}$-state (called "seeds") and a subsequent pulse performs the excitation from the ground state into the $\ket{\uparrow}$ state.
The excitation to the spin-down state $\ket{\downarrow} = \ket{51\mathrm{S}_{1/2}, m_J=1/2}$ is realized by a $\SI{1}{\mu s}$ long resonant two-photon excitation pulse.
The number and the spatial distribution of seed excitations created this way are controlled by the coupling strength $\Omega_\mathrm{S}$ and the Rydberg blockade conditions \cite{Lukin2001, Tong2004}.
After a variable delay time $\tau$, we apply the $\SI{1}{\mu s}$ long probe pulse by weakly driving ($\Omega_\mathrm{P} \ll \Omega_\mathrm{S}$) a single photon transition from the $\ket{5\mathrm{S}_{1/2}}$ ground state to the $\ket{\uparrow} = \ket{51\mathrm{P}_{3/2}, m_J=1/2}$ Rydberg state.
The spontaneous decay of Rydberg atoms into ions allows for continuous and time-resolved probing of the Rydberg population. For further experimental details see Methods.
Without probe pulse, the seed excitations decay on a typical timescale of $\tau_\mathrm{eff}\lesssim\SI{15}{\mu s}$.
The delay between the pump and the probe pulse is either chosen to be $\tau=\SI{1}{\mu s} \ll \tau_\mathrm{eff}$ to create an $\ket{\uparrow}$-excitation in the presence of the $\ket{\downarrow}$-seeds (interacting case) or it is chosen to be $\tau=\SI{300}{\mu s} \gg \tau_\mathrm{eff}$ to obtain a reference measurement of the temporally separated $\ket{\downarrow}$ and $\ket{\uparrow}$ excitations (non-interacting case).
By changing the probe laser detuning, we observe the spectroscopic response of the $\ket{\uparrow}$-excitation in the presence of a variable number of $\ket{\downarrow}$-excitations defined by $\Omega_\mathrm{S}$.
Effectively, this allows us to spectroscopically probe the random hopping model induced by the resonant dipole--dipole interaction between $\ket{\mathrm{S}}$- and $\ket{\mathrm{P}}$-states.

A series of spectra for increasing power of the pump laser pulse is shown in Fig.\,\ref{fig:Fig2}c-f together with a reference measurement in the absence of any seed excitation. One can clearly see that the spectroscopic line shape for the excitation of the $\ket{\uparrow}$-state significantly broadens.
While for small Rabi coupling, we see only little deviations from the non-interacting reference measurement, the line shape becomes largely modified for strong pump powers.
On the one hand, the $C_3$ Rydberg blockade manifests itself through the suppression of the spectroscopic signal on resonance.
On the other hand, the dipole--dipole-induced anti-blockade shows up as a strong enhancement of the signal far from resonance \cite{Ates2007, Amthor2010}.

In the reference spectra as well as for small pump power where we statistically see many realizations with no seed excitation, the line shape shows an increased signal for negative detunings which can be attributed to the formation of ultralong-range Rydberg molecules \cite{Bendkowsky2009}.
In the presence of seeds, however, the formation of these molecules becomes strongly suppressed due to the reduced probability to find an atom that simultaneously has the proper distance to the seeds and a ground state atom to form the molecule.

\section*{Spin model simulation}
In order to model our experimental spectra we perform Monte Carlo simulations of the random XY model Eq.\,\eqref{eq:Hamiltonian}.
We initialize an ensemble with random particle positions drawn from the BEC density distribution.
We restrict our treatment to the weak probing regime ($\Omega_\mathrm{P} \ll \Omega_\mathrm{S}$) and only consider the single-excitation subspace spanned by the states $\ket{i} = \ket{\mathrm{S}_1,\mathrm{S}_2,\dots\mathrm{P}_i\dots\mathrm{S}_{n}\mathrm{S}_{n+1}}$ where the single atom excited into the $\ket{\mathrm{P}}$-state is sitting on position $i$.
This subspace is simulated by choosing $n$ particles from the ensemble under blockade condition (representing the seed $\ket{\mathrm{S}}$-excitations) and an additional, randomly positioned particle (representing the $\ket{\mathrm{P}}$-excitation).
For each combination of the two free parameters, i.e. the number of seed excitations $n$ and the blockade radius $r_B$, the eigenvectors $\ket{\chi}$ and eigenvalues $E_\chi$ of the Hamiltonian Eq.\,\eqref{eq:Hamiltonian} are numerically determined for $10^5$ realizations.

The resulting eigenvalue spectrum of the Hamiltonian is projected onto $\ket{n+1}$ to obtain the simulated spectra $\Gamma|_{n,r_B}(\nu) = \sum_{E_\chi \approx h\nu}\left|\braket{n+1|\chi}\right|^2$ for fixed parameters $n$ and $r_B$.
We also account for the fluctuations in the initial number of seeds and the appearance of molecules for negative detuning in the absence of seeds (see Methods).
The simulated spectra are fitted to the measured spectra by varying the average number of seed excitations $\bar{n}$, the blockade radius $r_B$ and the amplitude $A$, using a least squares method.
The resulting fitted line shapes are shown in Fig.\,\ref{fig:Fig2}c-f.
A remarkable quantitative agreement between calculated and measured spectra is achieved.
Both effects, the suppression on resonance and the strong enhancement at large detunings, are recovered.
The model also correctly predicts the small but noticeable asymmetry towards negative detunings.
It can therefore not be attributed to the creation of molecules, as they are not included in the model (except to describe the influence of measurements without seeds in the weak pumping limit).

The asymmetry is remarkable as the binary interaction of a single $\ket{\mathrm{S}}$- with a single $\ket{\mathrm{P}}$-excitation is strictly symmetric.
However, beyond this binary regime, which has been studied previously \cite{Reinhard2008, PhysRevLett.99.073002, Park2011}, the eigenspectrum itself gives rise to an asymmetry due to correlations in the random hopping matrix elements \cite{Scholak2014}. This effect even prevails in the absence of the weak $C_6$ interaction.
As expected, the fitted $\bar{n}$ increases with the Rabi frequency of the pump pulse up to $\bar{n}=5.4$.
For the largest coupling, we therefore probe the simultaneous coherent interaction of the $\ket{\uparrow}$-state with up to five $\ket{\downarrow}$-spins.
Comparing the fitted number of seeds with an independent estimate based on the absolute number of detected ions agrees for small Rabi frequencies.
For the highest prepared seed densities, however, we see deviations that might originate from fast redistribution processes like l-changing collisions or Penning ionization that rapidly depopulate the $\ket{\downarrow}$-state during the $\SI{1}{\mu s}$ of delay time.
Comparing the measured spectra with a model that does not include the $C_6$ term in Eq.\,\eqref{eq:Hamiltonian}, we see that the van der Waals interaction only provides minor corrections to the spectral shape.

We have also repeated the measurement using a different fine structure state $\ket{\uparrow^\prime}=\ket{51\mathrm{P}_{1/2}, m_J=1/2}$.
We find the same level of quantitative agreement, suggesting that the microscopic details of the atomic level structure play a minor role and our system is adequately described by the effective two-level spin Hamiltonian Eq.\,\eqref{eq:Hamiltonian}.
Throughout all measurements, we consistently obtain a blockade radius $r_B=\SI{3.4}{\mu m}$ which fits well to the expected collective blockade radius.

\section*{Localization transition}
We now look in more detail at the structure of the involved eigenstates.
Since the position of the $\ket{\uparrow}$-state is randomly chosen without any distance constraints, it can possibly be very close to one of the seeds (but not to more than one, due to the blockade between the seeds), as sketched in Fig.\,\ref{fig:Fig1}b.
In that case, two eigenstates $\ket{\xi}_\pm \approx 1/\sqrt{2}(\ket{i}\pm \ket{j})$ exist at high absolute energy where the $\ket{\mathrm{P}}$-state is essentially localized on the closely separated pair $\{i,j\}$.
On the other hand, if the distance between all excitations is similar, all hopping energy contributions are small, leading to a set of low-energy, highly delocalized states (Fig.\,\ref{fig:Fig1}c).
Since the localized states exist predominantly at high absolute energies and the delocalized states exist at low energies, the random hopping model is predicted to show a delocalization--localization transition for increasing energy \cite{Sun2008,Scholak2014,Abumwis2020}.
This transition has also been connected to the appearance of a mobility edge \cite{Scholak2014}.

\begin{figure}
    \begin{center}
        \includegraphics[width=\columnwidth]{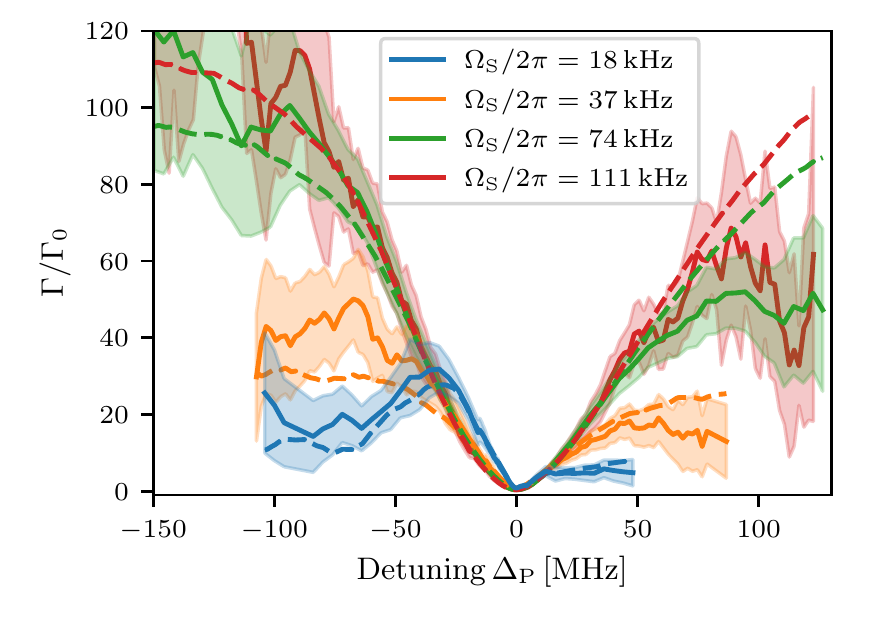}
    \end{center}
    \caption{Ratio $\Gamma(\Delta_\mathrm{P})/\Gamma_{0}(\Delta_\mathrm{P})$ of the interacting and non-interacting spectra.
    With increasing detuning, the ratio approaches a constant value on both sides of the resonance.
        A constant ratio indicates the occurrence of predominantly localized states.        
        For increasing Rabi frequency $\Omega_\mathrm{S}$ and thus increasing seed atom number the onset of this delocalization--localization transition shifts to larger detunings.
        The measured data (solid lines) is processed with a running average filter over 5 neighboring data points, the shaded areas correspond to errors obtained from error propagation of statistical errors of the interacting spectrum $\Gamma\left(\Delta_\mathrm{P}\right)$ and fitting errors of the non-interacting spectrum $\Gamma_{0}\left(\Delta_\mathrm{P}\right)$.
        The dashed lines represent the ratio $\bar{\Gamma}|_{\bar{n}, r_B}/\Gamma|_{0, r_B}$ of the simulated spectra in Fig.\,\ref{fig:Fig2}.}
    \label{fig:Fig3}
\end{figure}

Spectroscopically, the localization becomes visible in the tails of the spectral density $f(\Delta_\mathrm{P})$ where the signal of the localized pair-states decays algebraically $f(|\Delta_\mathrm{P}|\rightarrow \infty) \propto |\Delta_\mathrm{P}|^{-2}$ due to the $R^{-3}$-dependence of the interaction.
Thus, for large detunings, the interacting spectra $\Gamma\left(\Delta_\mathrm{P}\right)$ are expected to show the same $|\Delta_\mathrm{P}|^{-2}$ behavior as the Lorentzian line shape $\Gamma_0\left(\Delta_\mathrm{P}\right)$ of an isolated excitation.
Consequently, the ratio between the measured interacting spectrum and a Lorentzian fitted to the reference spectrum (only for $\Delta_\mathrm{P} > 0$ where no Rydberg molecules appear) is expected to become constant for large detunings.
These ratios, shown for different two-photon Rabi frequencies $\Omega_\mathrm{S}$ in Fig.\,\ref{fig:Fig3}, collapse on a common curve for small energies but start to deviate from this common behavior for larger detuning.
Starting from this point, the slope of the ratio decreases and is tending towards a constant value. This is the case for the two lowest seed numbers, thus indicating the presence of localized pair-states.
The energy at which the deviation appears is increasing with the number of seeds, in accordance with theoretical predictions \cite{Scholak2014}.

\begin{figure}
    \begin{center}
        \includegraphics[width=\columnwidth]{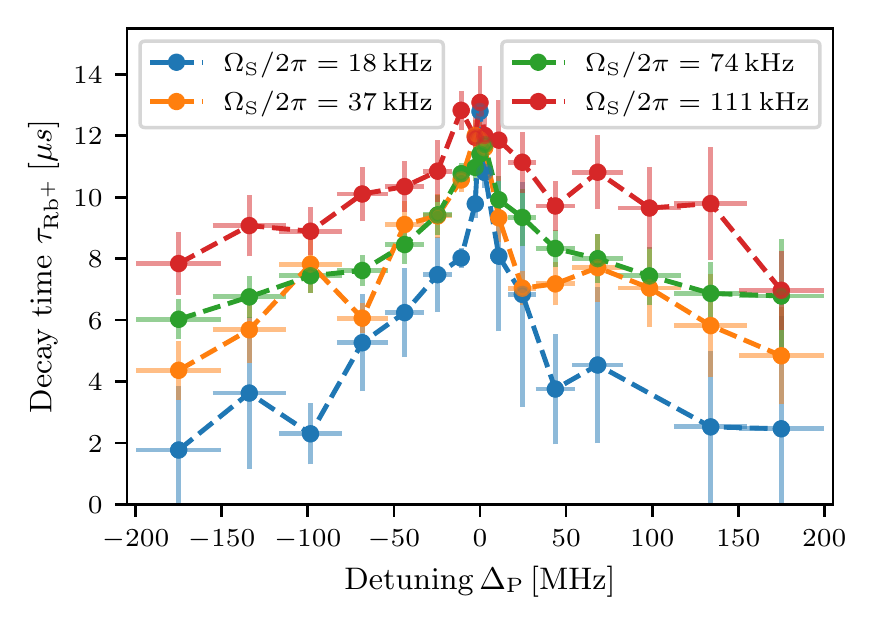}
    \end{center}
    \caption{Decay time $\tau_{\mathrm{Rb}^+}$ of the ionization channel $\mathrm{Rb}^+$.
        With increasing detuning $\Delta_\mathrm{P}$ the decay time drops.
        Increasing the Rabi frequency $\Omega_\mathrm{S}$ and thus increasing the seed atom number for a fixed detuning leads to a longer decay time, indicating a larger fraction of delocalized states.
        The decay times are obtained numerically from the ion signal under the assumption of an exponential decay of the $\mathrm{Rb}^+$ channel of the time-resolved spectra.
        Horizontal error bars reflect energy bin sizes for the evaluation. Decay time errors are given by statistical errors.}
    \label{fig:Fig4}
\end{figure}

Since high energetic dimer states are associated with small interparticle distances inside the $C_3$ potential, they are subject to strong acceleration and rapid motion.
The giant interaction cross section of the moving Rydberg atoms with the surrounding dense bath of ground state particles leads to efficient ionization \cite{Niederpruem2015}.
Due to the increasing fraction of pair-localized states, we thus expect the lifetime of the Rydberg excitations to decrease for increasing laser detuning.
To this end, we analyze the decay time of the ion signal after the probe pulse.
Fig.\,\ref{fig:Fig4} shows the extracted decay times $\tau_\mathrm{Rb^+}$ of the $\mathrm{Rb}^+$ ion signal that decrease with the laser detuning, signaling the increasing contribution of localized states.
For a higher number of seeds the delocalized states dominate over an increasingly large energy range.
Thus, the extracted lifetimes drop slower with energy as the pumping strength is increased.
Both experimental findings, the asymptotic $|\Delta_\mathrm{P}|^{-2}$-scaling of the spectra as well as the reduced lifetime of the Rydberg excitations with increasing detuning suggest that the system exhibits a localization--delocalization transition.

Interacting Rydberg gases are an almost perfect model system to study localization and transport phenomena as one can tune the relative strength of the hopping term and the long-range interaction by a proper choice of the involved Rydberg states.
The recent advent of tweezer arrays provides ideal conditions to look in detail at the emerging spatial structure of the localized states.
Thereby, the perfect control over each and every single atom even in 3D \cite{Barredo2018} allows the reproducible creation of tailored disordered patterns.
The inclusion of local excitation and readout processes would open up new prospects of probing the microscopic physics of transport dynamics in the context of open quantum systems. 

\section*{Methods}
\textbf{Experimental procedure.}
Starting from a 3D magneto-optical trap, we prepare a Bose-Einstein condensate of $\approx\num{90e3}$ $^{87}\mathrm{Rb}$ atoms, spin polarized in the $\ket{5\mathrm{S}_{1/2}, F=2, m_F=2}$ ground state, by performing forced evaporative cooling in a crossed YAG dipole trap with final trapping frequencies $\omega_t \approx 2 \pi \times \SI{160}{Hz}$ and $\omega_a \approx 2 \pi \times \SI{90}{Hz}$ in transverse and axial direction, respectively.
The $\ket{51\mathrm{S}_{1/2}, m_J=1/2}$ seed excitations are created with a two photon transition using a combination of continuous wave lasers at $\SI{420}{nm}$ and $\SI{1015}{nm}$.
Due to a large blue detuning $\Delta_\mathrm{int}=\SI{160}{MHz}$ to the intermediate $\ket{6\mathrm{P}_{3/2}}$ state, it can be adiabatically eliminated, allowing to describe the excitation with an effective Rabi frequency $\Omega_\mathrm{S}$.
The power of the infrared coupling laser is kept constant at $\SI{450}{mW}$ with a $1/e^2$ diameter of $\SI{150}{\mu m}$, the power of the weak blue beam ($1/e^2$ diameter of $\SI{1.7}{mm}$) is varied to set effective Rabi frequencies $\Omega_\mathrm{S}$ between $2\pi\times\SI{18}{kHz}$ and $2\pi\times\SI{111}{kHz}$.
The coupling of the ground state with the $\ket{51\mathrm{P}_{3/2}, m_J=1/2}$ state is generated with a frequency doubled continuous wave dye laser at $\SI{297}{nm}$ with a $1/e^2$ diameter of $\SI{100}{\mu m}$.
The Rabi frequency is fixed to $\Omega_\mathrm{P}\approx 2\pi\times\SI{4.5}{kHz}$.
Both pump $\ket{\mathrm{S}}$- and probe $\ket{\mathrm{P}}$-excitation pulses have a duration of $\SI{1}{\mu s}$.
The probe pulse either occurs at a delay $\tau=\SI{1}{\mu s}$ or $\tau=\SI{300}{\mu s}$ after the pump pulse, corresponding to the interacting and non-interacting case, respectively.
Using a small electric field ($E \approx \SI{50}{\mV\per\cm}$) we continuously guide the ions created from intrinsic ionization processes of the Rydberg atoms \cite{Niederpruem2015} to a discrete dynode detector.
This allows us to record a time resolved ion signal proportional to the Rydberg population.\\

\textbf{Simulations.}
For the numerical spin model simulation we randomly draw particle positions obeying a Thomas-Fermi distribution ($N=\num{90e3}$ and Thomas-Fermi radii $r_\mathrm{TF} = (4.6, 8.2, 4.6)\si{\mu m}$).
While, due to the complex ionization channels, the exact number of created $\ket{\uparrow}$-excitations is hard to determine precisely, we estimate it to be on the order of one.
Thus, we restrict our treatment to the weak probing regime ($\Omega_\mathrm{P} \ll \Omega_\mathrm{S}$) and only consider the single-excitation subspace spanned by the states $\ket{i} = \ket{\mathrm{S}_1,\mathrm{S}_2,\dots\mathrm{P}_i\dots\mathrm{S}_{n}\mathrm{S}_{n+1}}$ where the single $\ket{\mathrm{P}}$-excitation resides on position $i$.
Obviously, in the single-excitation subspace, the van der Waals interaction between $\ket{\mathrm{P}}$-states in Eq.\,\eqref{eq:Hamiltonian} vanishes.
This subspace is simulated by choosing $n$ particles from the ensemble under blockade condition (representing the seed $\ket{\mathrm{S}}$-excitations) and an additional, randomly positioned particle (representing the $\ket{\mathrm{P}}$-excitation).
While the dipole--dipole interaction $C_3 = d^2/(4\pi \epsilon_0)\left(1-3\cos^2(\theta)\right)$ is calculated from the dipole matrix element $d$, the van der Waals interaction coefficient ($C_6^\mathrm{\downarrow}$) is obtained by fitting to pair-state potentials from an exact diagonalization of the many-level system\cite{Weber2017}.

Since the Rabi coupling $\Omega_\mathrm{P}$ is much smaller than the interaction energy of the probed states, we directly couple to the eigenstates of the system.
However, due to the single photon excitation process we can only couple to the $\ket{\mathrm{P}}$-state fraction of the eigenstate at a particular atom.
Thus, the resulting eigenvalue spectrum of the Hamiltonian is projected onto $\ket{n+1}$ to obtain the simulated spectra $\Gamma|_{n,r_B}(\nu) = \sum_{E_\chi \approx h\nu}|\braket{n+1|\chi}|^2$ for fixed parameters $n$ and $r_B$.
Finally, the statistical nature of the seed excitation process provides a Poisson distributed number of seeds $n$ across multiple realizations $p(n)=\frac{\bar{n}^n e^{-\bar{n}}}{n!}$, with the average seed excitation number $\bar{n}$.
This is taken into account in the simulation by taking the Poisson weighted sum of the calculated spectra $\bar{\Gamma}|_{\bar{n}, r_B}(\nu) = \sum_{i=0}^\infty p(i)\Gamma|_{i,r_B}(\nu)$.
The summation is truncated at $i=20$ in our simulations.
The simulated spectra $\bar{\Gamma}|_{\bar{n}, r_B}(\nu)$ are fitted to the measured spectra by varying the free parameters $\bar{n}$, $r_B$ and an amplitude $A$ using a least squares method.
The $p(0)$ contribution of the Poisson distribution takes an exceptional role here because in absence of seed excitations Rydberg molecules have a strong influence onto the spectral shape.
Thus, $\Gamma|_{0,r_B}(\nu)$ is modeled with the experimentally obtained non-interacting spectrum instead of a Lorentzian line shape.

%

\section*{Data availability} 
The data that support the plots within this paper and other findings of this study are available from the corresponding author upon request.

\section*{Acknowledgements}
C.L., T.N. and H.O. acknowledge financial support by the Deutsche Forschungsgemeinschaft (DFG) within Project No. 277625399-TRR 185 (B2).

\section*{Author Contribution}
C.L., J.B. and P.M. performed the experiments. C.L., T.K., P.M. and T.N. analysed the data. C.L. and T.N. performed numerical simulations. C.L. prepared the manuscript. H.O. supervised the project. All authors contributed to the data interpretation and the manuscript.

\section*{Competing financial interests}
The authors declare no competing financial interests.

\end{document}